\documentclass[twocolumn,trackchanges]{aastex62}
\usepackage[section]{placeins}
\usepackage{longtable}
\usepackage{hyperref}
\usepackage{url}
\usepackage{amsmath}
\newcommand{\pslsn}{$P(\textrm{SLSN-I})$}

\usepackage{enumitem}
\usepackage{comment}
\usepackage{fontawesome}

\usepackage[hang,flushmargin]{footmisc} 
\usepackage{url}

\newcommand{\STScI}{\affiliation{Space Telescope Science Institute, 3700 San Martin Dr, Baltimore, MD 21218, USA}}

\newcommand{\CfA}{\affiliation{Center for Astrophysics \textbar{} Harvard \& Smithsonian, 60 Garden Street, Cambridge, MA 02138-1516, USA}}

\newcommand{\Birmingham}{\affiliation{Birmingham Institute for Gravitational Wave Astronomy and School of Physics and Astronomy, University of Birmingham, Birmingham B15 2TT, UK}}
\newcommand{\CIERA}{\affiliation{Center for Interdisciplinary Exploration and Research in Astrophysics and Department of Physics and Astronomy, \\Northwestern University, 1800 Sherman Ave., 8th Floor, Evanston, IL 60201, USA}}

\newcommand{\Steward}{\affiliation{Steward Observatory, University of Arizona, 933 North Cherry Avenue, Tucson, AZ 85721, USA}}
\newcommand{\PSUa}{\affiliation{Department of Astronomy \& Astrophysics, The Pennsylvania State University, University Park, PA 16802, USA}}
\newcommand{\PSUb}{\affiliation{Institute for Computational \& Data Sciences, The Pennsylvania State University, University Park, PA 16802, USA}}
\newcommand{\PSUc}{\affiliation{Institute for Gravitation and the Cosmos, The Pennsylvania State University, University Park, PA 16802, USA}}
\newcommand{\IAIFI}{\affiliation{The NSF AI Institute for Artificial Intelligence and Fundamental Interactions}}

\shorttitle{FLEET 2.0}
\shortauthors{Gomez et al.}

\begin{document}
	
	\title{The First Two Years of FLEET: an Active Search for Superluminous Supernovae}
	
	\correspondingauthor{Sebastian Gomez}
	\email{sgomez@stsci.edu}
	
	\author[0000-0001-6395-6702]{Sebastian Gomez}
	\STScI
	
	\author[0000-0002-9392-9681]{Edo Berger}
	\CfA\IAIFI
	
	\author[0000-0003-0526-2248]{Peter K. Blanchard}
	\CIERA
	
	\author[0000-0002-0832-2974]{Griffin Hosseinzadeh}
	\Steward
	
	\author[0000-0002-2555-3192]{Matt Nicholl}
	\Birmingham
	
	\author[0000-0002-1125-9187]{Daichi Hiramatsu}
	\CfA\IAIFI
	
	\author[0000-0002-5814-4061]{V. Ashley Villar}
	\PSUa\PSUb\PSUc
	
	\author[0000-0002-5723-8023]{Yao Yin}
	\CfA
	
	\begin{abstract}
		
		In November 2019 we began operating FLEET (Finding Luminous and Exotic Extragalactic Transients), a machine learning algorithm designed to photometrically identify Type I superluminous supernovae (SLSNe) in transient alert streams. Using FLEET, we spectroscopically classified 21 of the 50 SLSNe identified worldwide between November 2019 and January 2022. Based on our original algorithm, we anticipated that FLEET would achieve a purity of about 50\% for transients with a probability of being a SLSN, \pslsn$>0.5$; the true on-sky purity we obtained is closer to 80\%. Similarly, we anticipated FLEET could reach a completeness of about 30\%, and we indeed measure an upper limit on the completeness of $\approx 33$\%. Here, we present FLEET 2.0, an updated version of FLEET trained on 4,780 transients (almost 3 times more than in FLEET 1.0). FLEET 2.0 has a similar predicted purity to FLEET 1.0, but outperforms FLEET 1.0 in terms of completeness, which is now closer to $\approx 40$\% for transients with \pslsn$>0.5$. Additionally, we explore possible systematics that might arise from the use of FLEET for target selection. We find that the population of SLSNe recovered by FLEET is mostly indistinguishable from the overall SLSN population, in terms of physical and most observational parameters. We provide FLEET as an open source package on GitHub \href{https://github.com/gmzsebastian/FLEET}{\faGithub}.
		
	\end{abstract}
	
	
	\section{Introduction}\label{sec:intro}
	
	Type I superluminous supernovae (hereafter, SLSNe) can be up to 100 times more luminous than ``normal" Type Ic SNe (SNe Ic). They are identified by their high luminosity and blue spectra, and much like their normal luminosity SNe Ic counterparts, SLSNe are thought to be the end stage of massive stripped-envelope stars \citep{Chomiuk11,Quimby11}. To date, we know of $\sim 200$ spectroscopically classified SLSNe (Gomez et al., in prep, \citealt{Chen22}). Many open questions remain about the nature of SLSNe, what powers them, their environments, their progenitors, and their connection to other transients (e.g., \citealt{Lunnan14, Nicholl17_mosfit, Blanchard20, Orum20, Yan20, Hosseinzadeh21, Gomez22}). To address these questions it is imperative we find SLSNe early and efficiently.
	
	Telescope resources worldwide can only spectroscopically classify $\sim 10$\% of all transients discovered by current optical time-domain surveys. Since SLSNe only make up $\sim 1.5$\% of all discovered transients \citep{Perley20_BTS}, we need efficient methods to select the most likely SLSNe candidates for spectroscopic follow-up and make the most efficient use of limited telescope time. Towards this end, we developed FLEET (Finding Luminous and Exotic Extragalactic Transients), a machine learning algorithm that can determine the probability of being a SLSN, \pslsn, for any transient, calculated using the properties of both its light curve and host galaxy \citep{Gomez20}. We presented three versions of FLEET, a rapid version trained on the first $20$ days of photometry, a late-time version that uses the first $70$ days of photometry, and a redshift version, the only one that includes the redshift as a required parameter. The rapid version is meant to be used for real-time classification, while the late-time version can provide more robust predictions and be used to create photometric samples for population studies. The redshift version is the most accurate of the three, but requires knowing the redshift of the transient a priori.
	
	The structure of this paper is as follows: in \S\ref{sec:twoyears} we present the results from our first two years of FLEET operations and confirm we have been able to accurately and efficiently discover SLSNe. In \S\ref{sec:validation} we describe the expanded training set used for FLEET and present updated performance metrics of purity and completeness, and explore how FLEET can be used to generate photometric samples of SLSNe. In \S\ref{sec:selection} we compare the SLSNe from FLEET to the overall sample of SLSNe and explore possible selection effects from FLEET, both in terms of physical and observational parameters. Finally, we summarize our conclusions in \S\ref{sec:conclusion}. FLEET is provided as a Python package on Github\footnote{\url{https://github.com/gmzsebastian/FLEET}} and Zenodo \citep{Gomez20_FLEET}, as well as included in the Python Package Index under the name {\tt fleet-pipe}.
	
	\startlongtable
	\begin{deluxetable*}{cccc|cccc}
		\tablecaption{Spectroscopically Classified SLSNe \label{tab:slsne}}
		\tablewidth{0pt}
		\tablehead{\multicolumn{8}{c}{SLSNe with \pslsn$\geq0.5$} \\ 
			\multicolumn{4}{c}{Classified by FLEET} & \multicolumn{4}{c}{Classified by Others} \\
			\hline
			\colhead{Name} & \colhead{Redshift} & \colhead{\pslsn} & \colhead{Reference} & \colhead{Name} & \colhead{Redshift} & \colhead{\pslsn} & \colhead{Reference}}
		\startdata
		SN\,2019zbv    &   0.370 & 0.77 & [1]     &    SN\,2019sgg    &   0.573 & 0.62 & [14]    \\
		SN\,2019zeu    &   0.390 & 0.87 & [1]     &    SN\,2020abjx   &   0.390 & 0.56 & [15]    \\
		SN\,2020abjc   &   0.219 & 0.51 & [2]     &    SN\,2020auv    &   0.250 & 0.83 & [16]    \\
		SN\,2020adkm   &   0.226 & 0.62 & [3]     &    SN\,2020rmv    &   0.270 & 0.64 & [17]    \\
		SN\,2020jii    &   0.396 & 0.84 & [4]     &                   &         &      &         \\
		SN\,2020myh    &   0.283 & 0.67 & [4]     &                   &         &      &         \\
		SN\,2020onb    &   0.153 & 0.85 & [4]     &                   &         &      &         \\
		SN\,2020xgd    &   0.454 & 0.86 & [5,6]   &                   &         &      &         \\
		SN\,2021ejo    &   0.440 & 0.65 & [7]     &                   &         &      &         \\
		SN\,2021gtr    &   0.303 & 0.47 & [8]     &                   &         &      &         \\
		SN\,2021hpc    &   0.240 & 0.63 & [9]     &                   &         &      &         \\
		SN\,2021txk    &   0.460 & 0.75 & [10]    &                   &         &      &         \\
		SN\,2021vuw    &   0.200 & 0.81 & [11]    &                   &         &      &         \\
		SN\,2021yrp    &   0.300 & 0.65 & [12]    &                   &         &      &         \\
		\hline
		\rule{0pt}{1ex} \\
		\multicolumn{8}{c}{SLSNe with \pslsn$<0.5$} \\ 
		\multicolumn{4}{c}{Classified by FLEET} & \multicolumn{4}{c}{Classified by Others} \\
		\hline
		\colhead{Name} & \colhead{Redshift} & \colhead{\pslsn} & \colhead{Reference} & \colhead{Name} & \colhead{Redshift} & \colhead{\pslsn} & \colhead{Reference} \\
		\hline
		SN\,2019itq    &   0.481 & 0.03 & [1]    & SN\,2019pud    &   0.114 & 0.16 & [18]    \\
		SN\,2019otl    &   0.514 & 0.11 & [1]    & SN\,2019szu    &   0.213 & 0.02 & [19]    \\
		SN\,2019pvs    &   0.167 & 0.03 & [1]    & SN\,2019unb    &   0.064 & 0.12 & [20]    \\
		SN\,2019sgh    &   0.344 & 0.06 & [1]    & SN\,2020aewh   &   0.345 & 0.06 & [21]    \\
		SN\,2019ujb    &   0.165 & 0.04 & [1]    & SN\,2020ank    &   0.249 & 0.28 & [22,23] \\
		SN\,2019xaq    &   0.200 & 0.02 & [1]    & SN\,2020exj    &   0.133 & 0.04 & [24]    \\
		SN\,2021xfu    &   0.320 & 0.06 & [13]   & SN\,2020qef    &   0.183 & 0.09 & [25]    \\
		&         &      &        & SN\,2020qlb    &   0.159 & 0.03 & [26]    \\
		&         &      &        & SN\,2020tcw    &   0.065 & 0.08 & [27]    \\
		&         &      &        & SN\,2020vpg    &   0.257 & 0.06 & [28]    \\
		&         &      &        & SN\,2020wfh    &   0.330 & 0.20 & [15]    \\
		&         &      &        & SN\,2020xga    &   0.440 & 0.11 & [29]    \\
		&         &      &        & SN\,2020znr    &   0.100 & 0.16 & [30]    \\
		&         &      &        & SN\,2020zzb    &   0.166 & 0.10 & [15]    \\
		&         &      &        & SN\,2021bnw    &   0.098 & 0.06 & [31]    \\
		&         &      &        & SN\,2021een    &   0.160 & 0.08 & [32]    \\
		&         &      &        & SN\,2021ek     &   0.193 & 0.13 & [33]    \\
		&         &      &        & SN\,2021fpl    &   0.115 & 0.11 & [34]    \\
		&         &      &        & SN\,2021hpx    &   0.213 & 0.06 & [35]    \\
		&         &      &        & SN\,2021kty    &   0.159 & 0.09 & [36]    \\
		&         &      &        & SN\,2021lwz    &   0.065 & 0.00 & [37]    \\
		&         &      &        & SN\,2021mkr    &   0.280 & 0.06 & [38,39] \\
		&         &      &        & SN\,2021nxq    &   0.150 & 0.03 & [40]    \\
		&         &      &        & SN\,2021rwz    &   0.190 & 0.24 & [41]    \\
		&         &      &        & SN\,2021ybf    &   0.130 & 0.07 & [42]    \\
		\enddata
		\tablecomments{List of all SLSNe spectroscopically classified between November 2019 and November 2021, either by FLEET or by other groups. We include the spectroscopic redshifts and \pslsn\ values from the late-time classifier, sorted alphabetically. Only the SLSNe with \pslsn$\geq0.5$ were used for measuring the purity. All SLSNe except the ones discovered by other groups with \pslsn$\geq0.5$ were used to estimate the completeness. 1: \cite{Gomez_2019zeu}; 2: \cite{Blanchard_2020abjc}; 3: \cite{Blanchard_2020adkm}; 4: \cite{Gomez_2020onb}; 5: \cite{Gomez_2020xgd}; 6: \cite{Weil_2020xgd}; 7: \cite{Gomez_2021ejo}; 8: \cite{Gomez_2021gtr}; 9: \cite{Gomez_2021hpc}; 10: \cite{Gomez_2021txk}; 11: \cite{Gomez_2021vuw}; 12: \cite{Gomez_2021yrp}; 13: \cite{Gomez_2021xfu}; 14: \cite{Yan_2019sgg}; 15: \cite{Yan_2020abjx}; 16: \cite{Yan_2020auv}; 17: \cite{Terreran_2020rmv}; 18: \cite{Fremling_2019pud}; 19: \cite{Dahiwale_2019szu}; 20: \cite{Dahiwale_2019unb}; 21: \cite{Yan_2020aewh}; 22: \cite{Poidevin_2020ank}; 23: \cite{Dahiwale_2020ank}; 24: \cite{Dahiwale20_2020exj}; 25: \cite{Terreran_2020qef}; 26: \cite{Perez_2020qlb}; 27: \cite{Perley_2020tcw}; 28: \cite{Terreran_2020vpg}; 29: \cite{Gromadzki_2020xga}; 30: \cite{Ihanec_2020znr}; 31: \cite{Magee_2021bnw}; 32: \cite{Dahiwale_2021een}; 33: \cite{Gillanders_2021ek}; 34: \cite{Deckers_2021fpl}; 35: \cite{Gonzalez_2021hpx}; 36: \cite{Yao_2021kty}; 37: \cite{Perley_2021lwz}; 38: \cite{Chu_2021mkr}; 39: \cite{Poidevin_2021mkr}; 40: \cite{Weil_2021nxq}; 41: \cite{Weil_2021rwz}; 42: \cite{Bruch_2021ybf}.}
	\end{deluxetable*}
	
	\section{FLEET 1.0: Operations and Discoveries}\label{sec:twoyears}
	
	We began operating FLEET in November 2019, with the aim of finding SLSNe in ongoing transient alert streams. We selected candidates from transients reported to the Transient Name Server (TNS)\footnote{\label{ref:tns}\url{https://www.wis-tns.org/}}, and followed up the ones that were most likely to be SLSNe using: the Blue Channel \citep{schmidt89} and Binospec \citep{Fabricant19} spectrographs on the MMT 6.5-m meter telescope; and the Low Dispersion Survey Spectrograph (LDSS3c; \citealt{stevenson16}), and Inamori-Magellan Areal Camera and Spectrograph (IMACS; \citealt{dressler11}) on the Magellan 6.5-m telescopes. Since we began running FLEET and up to January 2022, there have been 50 SLSNe spectroscopically classified worldwide, and FLEET is responsible for classifying 21 of them (42\%), the full list is shown in Table~\ref{tab:slsne}. For completeness, in Table~\ref{tab:not} we include a list of 17 transients classified by FLEET that turned out to be something other than SLSNe (6 SNe IIn, 6 SNe Ia, 3 SNe II, and 2 SLSN-II). Where we define a SLSN-II as a SN IIn with a peak absolute magnitude $M_r < -20$. The classifications and spectra of all SNe classified as part of this program were published on the TNS.
	
	Purity is defined as the number of true positive SLSNe divided by the sum of true and false positive SLSNe. For our purity measurements we only consider transients with \pslsn$\geq0.5$, whether they are SLSNe ($N = 14$) or not ($N = 4$). Completeness is defined as the number of true positive SLSNe divided by the sum of true positive SLSNe and false negative SLSNe. Since we do not know the true number of undiscovered SLSNe, we cannot measure the true completeness of the classifier. Nevertheless, we can obtain an upper limit on the completeness by considering all SLSNe discovered by FLEET with \pslsn$\geq0.5$ ($N = 14$) divided by the total number of SLSNe discovered worldwide ($N = 46$), whether classified by FLEET or by someone else in the community. We do not include the four SLSNe with \pslsn$\geq0.5$ discovered by others in our estimate of completeness, to be more conservative about our estimates. We find that the on-sky performance surpassed the expected performance for almost every metric:
	
	\begin{itemize}
		\setlength\itemsep{0.0in}
		\item Rapid Purity = 84.6\% (Predicted 50\%)
		\item Late-time Purity  = 77.8\% (Predicted 59\%)
		\item Rapid Completeness = 33.3\% (Predicted 30\%)
		\item Late-time Completeness  = 30.4\% (Predicted 44\%)
	\end{itemize}
	
	The on-sky measurements of purity are particularly successful, given that FLEET was originally designed to optimize for purity as opposed to completeness. Additionally, it is reassuring to see that even when optimizing for purity, the measured completeness is consistent with the predicted level. In addition to the SLSN candidates with high \pslsn\ values, we observed some transients which had \pslsn$<0.5$ when other methods independent of FLEET suggested they could be possible SLSNe. For example, we targeted SN\,2021jmm, SN\,2021jtu, and SN\,2021owc since they were part of the ZTF SLSN candidates reports \citep{Perley_2021jtu, Yan_2021jmm, Yan_2021owc}. We also observed a few targets as part of an experimental attempt to target Type-II SLSNe, which although they had low \pslsn\ values, their $P(\textrm{SLSN-II})$ values are above 0.7: SN\,2019itq, SN\,2019otl, SN\,2019pvs, SN\,2019sgh, SN\,2021xfu (all spectroscopically classified as Type-I SLSNe), SN\,2021owc, SN\,2021srg (SN IIn), and SN\,2020mad (SLSN-II). We manually vetted every transient before triggering spectroscopic follow-up, which means we did not observe every transient with \pslsn$>0.5$, which might have lowered our measured purity. 
	
	The redshift classifier was originally designed with the goal of using with data from the Legacy Survey of Space and Time (LSST), once it provides photometric redshift estimates for galaxies down to $i\approx 25$ mag \citep{Graham18}. Therefore, we do not include a comparable validation of the redshift classifier, since we have not yet tested it on active surveys. 
	
	\section{FLEET 2.0: Updates and Predictions}\label{sec:validation}

\begin{deluxetable}{ccccc}
	\tablecaption{Non-SLSNe Classified by FLEET \label{tab:not}}
	\tablewidth{0pt}
	\tablehead{
		\colhead{Name} & \colhead{Type} & \colhead{Redshift} & \colhead{\pslsn} & \colhead{Reference}    }
	\startdata
	\multicolumn{5}{c}{SNe with \pslsn$\geq0.5$}  \\ 
	\hline
	SN\,2020mad  & SLSN-II & 0.123 & 0.60 & [1]   \\
	SN\,2020tyk  & SN IIn  & 0.087 & 0.73 & [2,3] \\
	SN\,2021abjs & SN Ia   & 0.160 & 0.52 & [4]   \\
	SN\,2021ali  & SLSN-II & 0.192 & 0.80 & [6]   \\
	\hline
	\multicolumn{5}{c}{SNe with \pslsn$<0.5$}     \\ 
	\hline
	SN\,2020hvw  & SN II   & 0.093 & 0.41 & [1]   \\
	SN\,2020oqy  & SN Ia   & 0.135 & 0.12 & [1]   \\
	SN\,2021aeaj & SN Ia   & 0.205 & 0.47 & [5]   \\
	SN\,2021jmm  & SN IIn  & 0.190 & 0.45 & [4]   \\
	SN\,2021jtu  & SN IIn  & 0.113 & 0.48 & [4]   \\
	SN\,2021osr  & SN IIn  & 0.085 & 0.01 & [7,6] \\
	SN\,2021owc  & SN IIn  & 0.127 & 0.42 & [6]   \\
	SN\,2021rcn  & SN Ia   & 0.121 & 0.05 & [8]   \\
	SN\,2021srg  & SN IIn  & 0.069 & 0.07 & [5]   \\
	SN\,2021stu  & SN II   & 0.034 & 0.27 & [9]   \\
	SN\,2021uuz  & SN II   & 0.175 & 0.36 & [10]  \\
	SN\,2021wkg  & SN Ia   & 0.120 & 0.06 & [11]  \\
	SN\,2021zqa  & SN Ia   & 0.120 & 0.34 & [11]  \\
	\enddata
	\tablecomments{List of all transients spectroscopically classified by FLEET that turned out to not be SLSNe, including their redshift and \pslsn\ estimate from the late-time classifier, sorted alphabetically. Only the SNe with \pslsn$\geq0.5$ were used to calculate our purity estimates. 1: \cite{Gomez_2020onb}; 2: \cite{Gomez_2020xgd}; 3: \cite{Weil_2020xgd}; 4: \cite{Gomez_2021hpc}; 5: \cite{Gomez_2021txk}; 4: \cite{Gomez_2021abjs}; 5: \cite{Gomez_2021aeaj}; 6: \cite{Hosseinzadeh_2021ali}; 7: \cite{Chu_2021osr}; 8: \cite{Gomez_2021txk}; 9: \cite{Gomez_2021stu}; 10: \cite{Gomez_2021uuz}; 11: \cite{Gomez_2021wkg}.}
\end{deluxetable}

	The original version of FLEET was trained on 1,813 spectroscopically classified transients. Here, we re-train FLEET on an updated and more extensive list of 4,780 spectroscopically classified transients with the following distinct labels from the TNS: 2,983 SN\,Ia, 749 SN\,II, 187 SLSN-I, 157 SN\,IIn, 143 CV, 105 SN\,Ic, 89 SN\,IIP, 80 SN\,Ib, 68 SN\,IIb, 52 SLSN-II, 46 TDE, 35 SN\,Ic-BL, 26 SN\,Ibc, 23 AGN, 19 SN\,Ibn, and 18 variable star. We re-train the algorithm using the same procedures outlined in \cite{Gomez20}, where we optimize the depth of the random forest trees, the features included, and the number of days of photometry to consider. We find that the optimal parameters we determined for FLEET 1.0 still perform best for FLEET 2.0, with the exception that a tree depth of 11 branches now results in a higher completeness than the original depth of 7 branches.
	
\begin{figure}
		\begin{center}
			\centering
			{\includegraphics[width=\columnwidth]{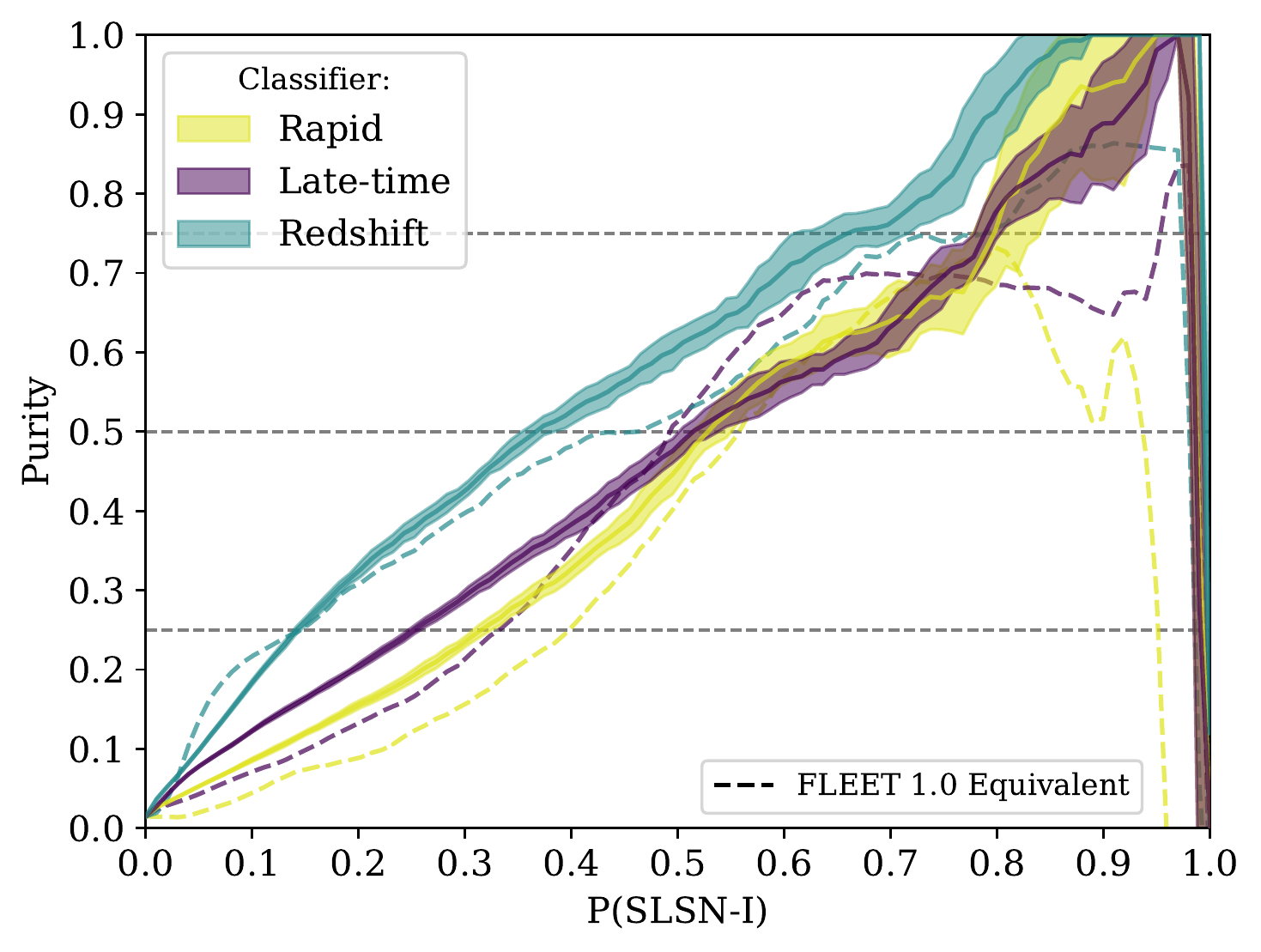}}
			{\includegraphics[width=\columnwidth]{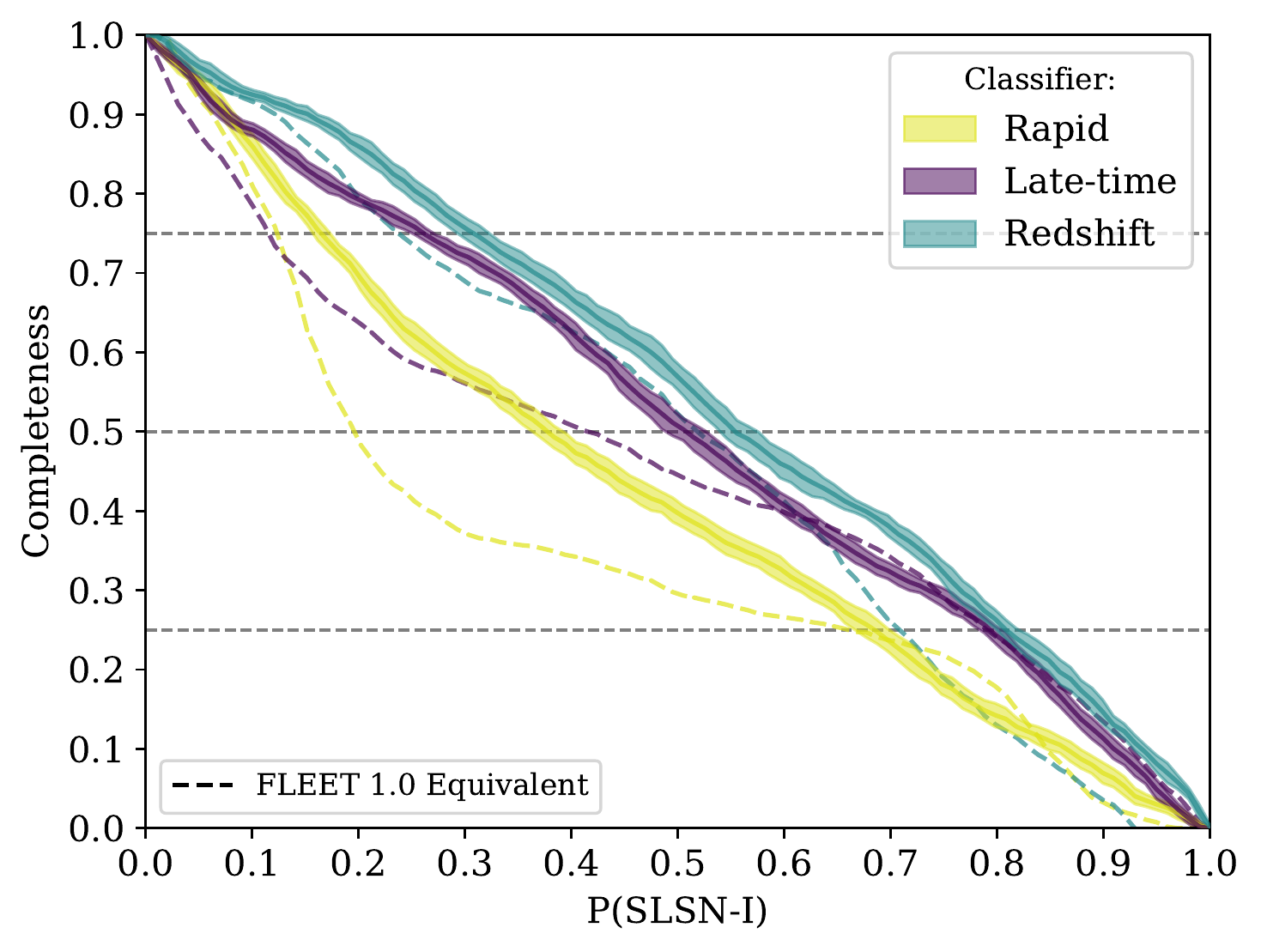}}
			\caption{Purity and completeness estimates for FLEET 2.0. \textit{Top}: Purity as a function of classification confidence for the rapid, late-time, and redshift versions of the classifier. \textit{Bottom}: completeness as a function of classification confidence for the same classifiers. In dashed lines, we show the equivalent measurements from FLEET 1.0. \label{fig:purity}}
		\end{center}
	\end{figure}

	In Figure~\ref{fig:purity} we show the updated purity and completeness obtained from the new FLEET 2.0, compared to those of FLEET 1.0. We find that for transients with \pslsn$\geq0.5$ the expected purity is $\approx 50$\% for both the rapid and late-time classifiers. The expected completeness for the same threshold is $\approx 50$\% for the late-time classifier and $\approx 40$\% for the rapid classifier. FLEET 1.0 had a comparable purity to FLEET 2.0, but a completeness $\approx 10$\% lower for both classifiers. For the redshift classifier, the purity increased in FLEET 2.0 compared to FLEET 1.0, from $\approx 50$\% to $\approx 60$\% and the completeness increased from $\approx 55$\% to $\approx 60$\%. 
	
	\begin{figure}
		\begin{center}
			\centering
			{\includegraphics[width=\columnwidth]{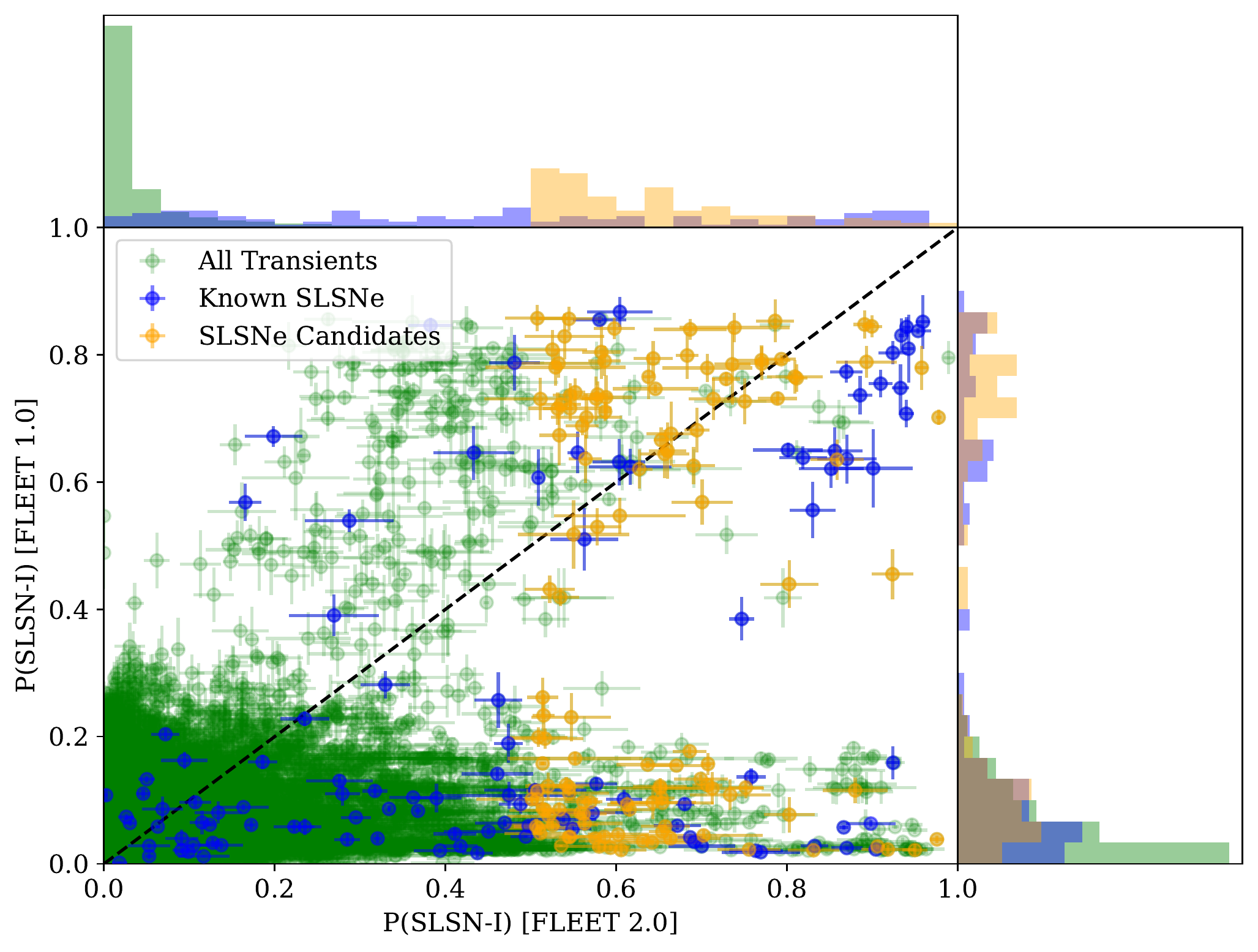}}
			\caption{Classification confidence \pslsn\ for both the new and old versions of FLEET, using the late-time classifier. We find that some SLSNe that had low values of \pslsn\ in FLEET 1.0 are now well above 0.5 in FLEET 2.0.
				\label{fig:new}}
		\end{center}
	\end{figure}
	
	\subsection{Photometric Sample of SLSNe}\label{sec:missing}
	
	We use FLEET 2.0 to generate a photometric sample of SLSN candidates by running it on every optical transient reported to the TNS and searching for unclassified SLSNe. Since these are mostly transients that have faded away by the time of this analysis, we use the late-time classifier that was trained on the first 70 days of photometry. For this reason, we exclude transients that were too young, discovered after 1 March 2022. At that time, there were a total of 95,729 transients reported to the TNS. Of these, 83,589 have a Decl. $> -32$ deg, required for FLEET to query the PS1/$3\pi$ \citep{Chambers18} or Sloan Digital Sky Survey (SDSS) catalogs \citep{Alam15,Ahumada19}. Of these, 34,805 were discovered before ZTF began operations, and have either only one data point or minimal discovery photometry available. An additional $2,078$ transients were excluded because they only had coverage in one band, despite having tens to hundreds of detections in ZTF. Of the remaining transients, 31,922 had at least 2 $g$-band and 2 $r$-band points, required for FLEET to fit a model to their light curve. Finally, we excluded 30 transients that were located in regions with a Galactic extinction $\gtrsim 1.5$ mag. The final list of transients on which we run FLEET contains 31,892 transients.
	
	In Figure~\ref{fig:new} we show the \pslsn\ values from FLEET 2.0 compared to the estimates from FLEET 1.0 for all 31,892 transients. The increase in completeness of FLEET 2.0 is evident in the cluster of transients that had low \pslsn\ values in FLEET 1.0 that now lie at \pslsn$>0.5$. In Figure~\ref{fig:new}, we also include the sample of 106 spectroscopically classified SLSNe, shown in blue. A total of 69 of these SLSNe had \pslsn$<0.2$ in FLEET 1.0, 20 of which now have \pslsn$>0.5$ in FLEET 2.0, representing the improvement in completeness. While 21 SLSNe now have a lower \pslsn\ in FLEET 2.0, 85 have a higher \pslsn\ in FLEET 2.0 compared to FLEET 1.0, indicating a net gain. Of the 21 SLSNe that lowered in confidence, only 6 decreased significantly from \pslsn$> 0.5$ in FLEET 1.0 to \pslsn$<0.5$ in FLEET 2.0. 
	
	There were 25 SLSNe classified by other groups, which had \pslsn$<0.5$ in FLEET 1.0. Of these, 9 have \pslsn$>0.5$ in FLEET 2.0. Of the remaining ones, 9 still have significantly low values of \pslsn$<0.25$. These low \pslsn\ values are mostly due to sparse or noisy light curves, and in the case of SN\,2021lwz, a very quickly evolving light curve.
	
	Of the 106 spectroscopically confirmed SLSNe in the sample shown in Figure~\ref{fig:new}, 45 have \pslsn$>0.5$, corresponding to a completeness of $43$\%. We find a total of 382 transients with \pslsn$>0.5$, including the 45 known SLSNe. This corresponds to a purity of $12$\%, much lower than the on-sky purity presented in \S\ref{sec:twoyears} for two main reasons. First, obvious failure modes still exist that pollute the sample (e.g., variable AGN, or variable stars with long light curves and no detected host can be misclassified as SLSNe). We can visually purge these obvious failure modes to increase the purity of our sample. Second, there are likely still some SLSNe in those 382 transients that have not been classified, but that can be considered to create a photometrically selected sample of SLSNe. Of those 382 transients, we exclude 139 from being possible SLSNe because either their light curves are too noisy or sparse, or they strongly resemble a stellar transient or AGN (these are the green points in Figure~\ref{fig:new} with \pslsn$>0.5$ in FLEET 2.0). That leaves 243 transients as possible SLSNe, noting that none of the 45 spectroscopic classified SLSNe were rejected in the manual purging process. Of these, 70 transients are already classified as: 2 AGN, 2 CV, 11 SLSN-II, 1 SNI, 13 SNII, 2 SNIIP, 17 SNIIn, 16 SNIa, 2 SNIa-91T-like, 1 SNIa-CSM, 2 SNIc, and 1 TDE. Finally, 128 remain unclassified SLSN candidates. We show the 128 photometrically selected SLSNe candidates in orange in Figure~\ref{fig:new}. We find that 45\% of these candidates also had \pslsn$>0.5$ in FLEET 1.0, whereas the other half was previously below this threshold. If we were to spectroscopically classify these 128 SLSNe candidates, given their \pslsn, we expect about 82 of them would be true SLSNe, or a $\sim 40$\% increase of the current total population size of SLSNe. A full exploration of this population will be presented in future work.
	
	\section{Selection Effects}\label{sec:selection}
	
	\begin{figure*}
		\begin{center}
			\centering
			{\includegraphics[width=0.9\textwidth]{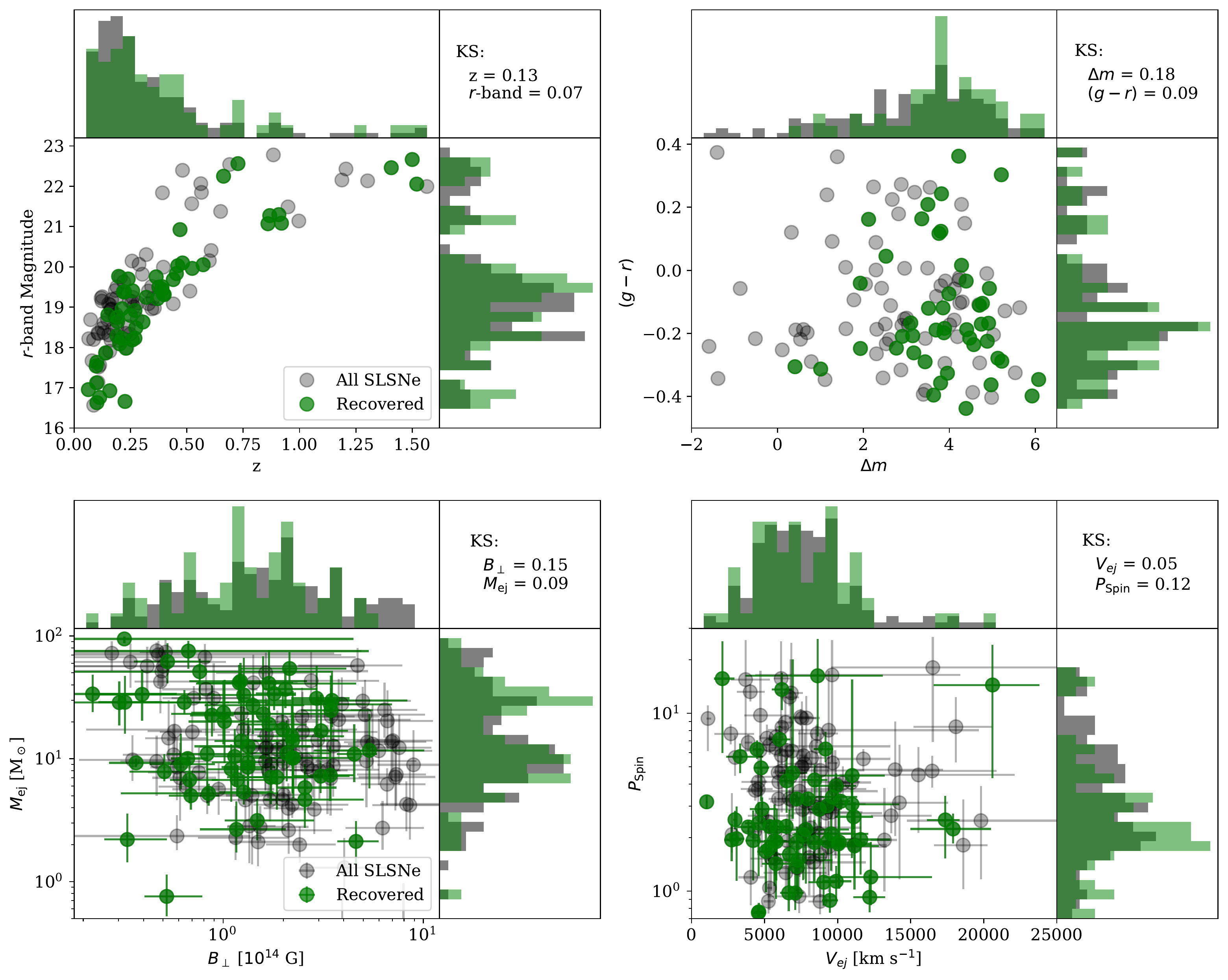}}
			\caption{The gray data points show the parameters of the full sample of SLSNe, and the green points are SLSNe that were recovered by FLEET using the late-time classifier with \pslsn$> 0.5$. \textit{Top Left}: ejecta mass and neutron star magnetic field. \textit{Top Right}: Magnetar spin period and ejecta velocity. \textit{Bottom Left}: Peak apparent $r$-band magnitude and redshift. {Bottom Right}: The $(g-r)$ color during peak and the host minus transient magnitude, $\Delta m$. We include the KS metric for each parameter on the upper right corner of each sub-plot. \label{fig:bias}}
		\end{center}
	\end{figure*}
	
	In this section, we explore whether the use of FLEET introduces biases or selection effects in terms of observed or physical SLSN parameters. For this test, we use the sample of 187 SLSNe used for training FLEET. We run these SLSNe through the late-time classifier in FLEET and compare the observational parameters of the SLSN population at large, with those of 74 ``recovered" SLSN with \pslsn$>0.5$. To quantify how different the parameter distributions are, we implement a two-sample Kolmogorov-Smirnov (KS), where a KS metric of $D = 0.0$ indicates the two samples are drawn from the same distribution, and $D = 1.0$ means there is no overlap between the distributions. We show how the apparent $r$-band magnitude, redshift, $(g-r)$ color, and $\Delta m$ compare between the two samples on the top panels of Figure~\ref{fig:bias}; where $\Delta m$ is defined as the host magnitude minus the transient peak magnitude. We find no obvious selection effects in terms of redshift $r$-band magnitude, or $(g-r)$ color, which have KS metrics (and $p$ values) of $0.13 (0.4)$, $0.07 (0.96)$, and $0.09 (0.3)$, respectively. The only observational parameter for which we do find a possible difference between the full and recovered sample is $\Delta m$, where SLSNe with $\Delta m < 0$ are generally not selected as likely SLSNe. This is reflected in its KS metric of $0.18 (0.09)$. To calculate $\Delta m$ here we use as reference the $r$-band magnitudes of the host galaxies obtained from either PS1/$3\pi$, SDSS, or from values published in compilation studies of the hosts of SLSNe \citep{Inserra13,Lunnan14,Angus16,Perley16,Schulze18,Angus19,Schulze21}, or studies of individual SLSNe \citep{Leloudas12,Vreeswijk14,Lunnan16,Yan17,Blanchard18,Lunnan18_iPTF16eh,Blanchard21,Yin22}.
	
	We also explore selection effects in terms of the physical parameters of the SLSN population. We derive physical parameters of the 187 SLSNe from light curves fits using {\tt MOSFiT}, a python code designed to model the light curves of transients using a variety of different power sources and derive their physical parameters \citep{guillochon18}. We show how the population of recovered SLSNe compares to the overall population in the bottom panels of Figure~\ref{fig:bias}. We find a KS metric (and $p$ value) of $0.15 (0.26)$ for the magnetic field $B_\perp$, $0.09 (0.79)$ for the ejecta mass $M_{\rm ej}$, $0.05 (0.99)$ for the ejecta velocity $V_{\rm ej}$, and $0.12 (0.03)$ for the spin period $P_{\rm Spin}$. This shows that the two populations are very similar, at the $5-10$\% level, and that FLEET is not introducing a bias in terms of physical parameters.
	
	In conclusion, we find that even if FLEET is less likely to select SLSNe with low values of $\Delta m$, this does not translate into a bias on derived physical parameters.
	
	\section{Conclusion}\label{sec:conclusion}
	
	We have provided an evaluation of the on-sky performance of FLEET during our first two years of operations. We include the list of 21 SLSNe found by FLEET 1.0 from November 2019 to January 2022, which represent 42\% of all SLSNe discovered worldwide in the same time period. We show that our estimate for purity of $\approx 50$\% from FLEET 1.0 was surpassed with a measured purity of $\approx 80$\%. The measured completeness was in line with our predicted value of $\approx 30$\%.
	
	We describe an updated version, FLEET 2.0, trained on 4,780 transients, or almost three times as many as FLEET 1.0. The updated estimate for purity and completeness are both $\approx 50$\% for the late-time classifier, and a $\approx 50$\% purity and $\approx 40$\% completeness for the rapid classifier. The estimated purity and completeness for the redshift classifier are both $\approx 60$\%. It is of course possible the on-sky purity will be higher, as was the case for FLEET 1.0. We test whether FLEET might be introducing selection effects in the recovered SLSN population, and find no significant biases against any physical or most observational parameters.
	
	Finally, we generate a photometric sample of 128 likely SLSN candidates selected by FLEET from ZTF archival data. We find that if we were to spectroscopically classify all of these candidates, given their \pslsn\ values, $\sim 80$ of them would result in true SLSNe, which would represent a $\sim 40$\% increase in the current population size of SLSNe.
	
	\acknowledgements
	
	S.G. is supported by an STScI Postdoctoral Fellowship. The Berger Time Domain group at Harvard is supported in part by NSF and NASA grants, including support by the NSF under grant AST-2108531, as well as by the NSF under Cooperative Agreement PHY-2019786 (The NSF AI Institute for Artificial Intelligence and Fundamental Interactions \url{http://iafi.org/}). M.N. is supported by the European Research Council (ERC) under the European Union’s Horizon 2020 research and innovation programme (grant agreement No.~948381) and by a Fellowship from the Alan Turing Institute. This research has made use of NASA’s Astrophysics Data System. This research has made use of the SIMBAD database, operated at CDS, Strasbourg, France. Operation of the Pan-STARRS1 telescope is supported by the National Aeronautics and Space Administration under grant No. NNX12AR65G and grant No. NNX14AM74G issued through the NEO Observation Program.
	
	\facilities{ADS, TNS}
	\software{Astropy \citep{astropy}, extinction \citep{Barbary16}, Matplotlib \citep{matplotlib}, emcee\citep{foreman13}, NumPy \citep{numpy}, scikit-learn \citep{Pedregosa12}, SMOTE \citep{Chawla02}, FLEET \citep{Gomez20_FLEET}}
	
	\bibliography{references}
	
\end{document}